\RequirePackage{fix-cm}
\documentclass[
 aps,prx,
 amsmath,amssymb,superscriptaddress,
 preprint
]{revtex4-1}

\usepackage{graphicx}
\usepackage{dcolumn}
\usepackage{bm}
\usepackage{color}
\usepackage{subfigure}
\usepackage[utf8]{inputenc}
\usepackage[T1]{fontenc}
\usepackage{mathptmx}
\usepackage{hyperref} 
\usepackage{makecell}
\usepackage[normalem]{ulem}
\usepackage{multirow}

\begin{document}

%\title{Multi-configurational nature of quantum states in diamond}
\title{Multi-configurational nature of electron correlation within nitrogen vacancy centers in diamond}

\author{Yilin Chen}
%\email{yilinchenpku@pku.edu.cn}
\affiliation{International Center for Quantum Materials, School of Physics, Peking University, Beijing 100871, P. R. China}

\author{Tonghuan Jiang}
%\email{jiangth1997@pku.edu.cn}
\affiliation{School of Physics, Peking University, Beijing 100871, P. R. China}

\author{Haoxiang Chen}
%\email{hxchen@pku.edu.cn}
\affiliation{School of Physics, Peking University, Beijing 100871, P. R. China}

\author{Erxun Han}
%\email{exhan@stu.pku.edu.cn}
\affiliation{School of Physics, Peking University, Beijing 100871, P. R. China}

\author{Ali Alavi}
%\email{a.alavi@fkf.mpg.de}
\affiliation{Max Planck Institute for Solid State Research, Heisenbergstrasse 1, 70569 Stuttgart, Germany}
\affiliation{University of Cambridge, Lensfield Road, Cambridge CB2 1EW, United Kingdom}

\author{Kuang Yu}
\email{yu.kuang@sz.tsinghua.edu.cn}
\affiliation{Tsinghua-Berkeley Shenzhen Institute (TBSI), Institute of Materials Research (iMR), Tsinghua Shenzhen International Graduate School (TSIGS), Tsinghua
University, Shenzhen, Guangdong Province 518055, P. R. China}

\author{Enge Wang}
\email{egwang@pku.edu.cn}
\affiliation{International Center for Quantum Materials, School of Physics, Peking University, Beijing 100871, P. R. China}
\affiliation
{Interdisciplinary Institute of Light-Element Quantum Materials and Research Center for Light-Element Advanced Materials, Peking University, Beijing 100871, P. R. China}
\affiliation{
 Collaborative Innovation Center of Quantum Matter, Beijing 100871, P. R. China
}
\affiliation{Songshan Lake Materials Lab, Institute of Physics, Chinese Academy of Sciences, Guangdong, P. R. China.}

\author{Ji Chen}
\email{ji.chen@pku.edu.cn}
\affiliation{School of Physics, Peking University, Beijing 100871, P. R. China}
\affiliation
{Interdisciplinary Institute of Light-Element Quantum Materials and Research Center for Light-Element Advanced Materials, Peking University, Beijing 100871, P. R. China}
\affiliation{
 Collaborative Innovation Center of Quantum Matter, Beijing 100871, P. R. China
}
\affiliation{Frontiers
Science Center for Nano-Optoelectronics, Peking University, Beijing 100871, P. R. China}

\date{\today}% It is always \today, today

\begin{abstract}
Diamond is a solid-state platform to develop quantum technologies, but it has been a long-standing problem that the current understanding of quantum states in diamond is mostly limited to single-electron pictures.
Here, we combine the full configuration interaction quantum Monte Carlo method and the density-matrix functional embedding theory, to achieve unprecedented accuracy in describing the many-body quantum states of nitrogen vacancy (NV) centers in diamond. 
More than 30 electrons and 130 molecular orbitals are correlated, which reveals the multi-configurational wavefunction of the many-body quantum states in diamond.
The multi-configurational description explains
puzzling experimental measurements in intersystem crossing and charge state transition in NV centers in diamond.
The calculations not only reproduce the available experimental measurements of the energy gaps between quantum states but also provide new benchmarks for states that are still subject to considerable uncertainty.
This study highlights the importance of multi-configurational wavefunction in the many-body quantum states in solids.
\end{abstract}

\maketitle

\section{Introduction}

Diamond has been a key component in many quantum technologies such as qubits, photon emitters, and nano-sensors,
utilizing the isolated and manipulable properties of the quantum states \cite{weber_quantum_2010,dreyer_first_2018}.
The nitrogen vacancy (NV) centers are systems having such ideal properties and various applications have already been demonstrated in experiments \cite{dolde_room-temperature_2013,waldherr_quantum_2014,liu_demonstration_2015,zaiser_enhancing_2016}.
Although there have been many detailed experimental and theoretical studies on quantum states in diamond from different perspectives, questions remain on the details of the low-lying spin states and transitions between charge states \cite{gali_ab}.
The mechanism of intersystem crossing between triplets and dark singlets in negatively charged NV center (NV$^{-}$), a process that is essential for initialization and readout of quantum states, is still controversial \cite{Doherty_2011,choi_mechanism}.
A better understanding of the ionization processes of NV centers to the neutral (NV$^{0}$) and the positively charged (NV$^{+}$) states are demanded because the latter states have demonstrated promising applicability in quantum devices \cite{pfender_protecting_2017,pezzagna_quantum_2021,zheng_coherence_2022}. 

Theoretically, the main challenge is to describe the many-electron states in the quantum regime from \textit{ab initio} calculations \cite{gali_ab}. 
The atomic model for an NV center would consist of at least three carbon atoms, one nitrogen atom and one saturating shell of hydrogen atoms, leading to a demand to correlate more than 30 electrons in hundreds of orbitals.
This means an exact description involves a gigantic number of Slater determinants, i.e. configurations, which are inherently correlated and should be considered on the same foot.
Thus far, the most widely accepted physical pictures of the electronic structure of NV centers are often based on mean-field theories, such as Hartree Fock and density functional theory (DFT) \cite{gali_kaxiras_2008,larsson_electronic_2008},
where only a single configuration is used to approximate the many-electron wavefunction \cite{booth_towards_2013}. 
Much of the understanding of the NV centers are deeply rooted in such a single-configuration picture, which has led to successful applications aforementioned.
However, in the last few years, deepened theoretical analyses and interpretations of experimental measurements have suggested that correlation effects in diamond are much stronger than anticipated, hence a multi-configurational description is urgently needed to go beyond the current understanding \cite{ma2010excited,ranjbar2011many,delaney2010spin,bockstedte2018ab,ma2020quantum}. 
%

% Flow chart and general strategy
\begin{figure}[htbp]
    \centering
    \includegraphics[width=12cm]{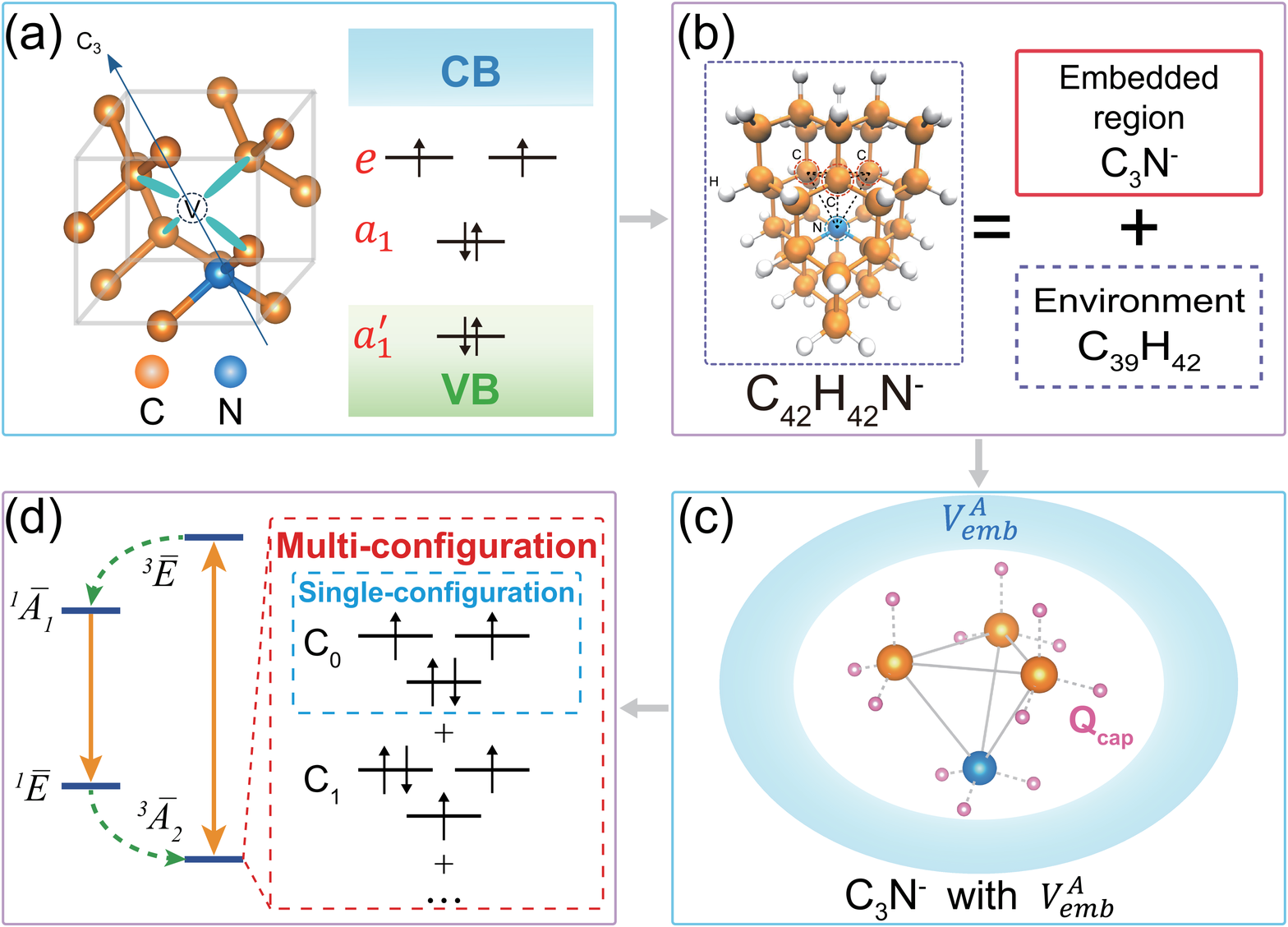}
    \caption{\textbf{Flow chart of this work.} (\textbf{a}) Left: the local structure of the NV$^{-}$ center in diamond. Right: a sketch of the single particle band structure and related states between the valence (VB) and conduction (CB) bands (degenerate $e_{x,y}$, $a_{1}$ and $a'_{1}$), where the arrows indicate the spin occupations of the ground state. (\textbf{b}) The partitioning scheme of a cluster C$_{42}$H$_{42}$N$^{-}$ containing the NV$^{-}$ center. (\textbf{c}) A sketch of constructing the DFT embedding potential using DMFET, orbital orthogonalization and projection. 
    (\textbf{d}) The multi-configurational electronic structure of the NV$^{-}$ center. Left: showing four important low-lying states: the ground triplet state $^{3}\bar{A}_{2}$, the optical excited triplet state $^{3}\bar{E}$ and the dark intermediate singlet states $^{1}\bar{A}_{1}$ and $^{1}\bar{E}$. The green dashed line highlight the intersystem crossing (ISC) processes $^{3}\bar{E}\leftrightarrow{^{1}\bar{A}_{1}}$ and $^{1}\bar{E}\leftrightarrow{^{3}\bar{A}_{2}}$. Right: the configurations of the ground state $^{3}\bar{A}_{2}$, the blue box represents the single-configurational picture while the red box highlights the strongly-correlated multi-configurational picture obtained by FCIQMC-in-DFT. 
    }
    \label{Fig_flowchart}
\end{figure}

% Our works' main approach
In this work, we develop a new computational framework, which allows us to perform full configuration interaction quantum Monte Carlo (FCIQMC) together with a quantum embedding technique, namely the density-matrix functional embedding theory (DMFET).
FCIQMC is a stochastic sampling method that solves the full configuration interaction problem, yielding accurate solutions to the quantum chemical hamiltonian in a large space of configurations \cite{booth_fermion_2009,Kai_NECI_2020,bogdanov_enhancement_2022}.
DMFET is a robust quantum embedding scheme accounting of the correlation effects from the outer regimes of the material quantum mechanically \cite{yu2017extending,zhang2018subspace}.
Our approach, dubbed as FCIQMC-in-DFT, can effectively overcome the difficulties encountered in previous theoretical studies, namely the simultaneous inclusion of an accurate quantum environment and full correlation at the site of the defect.
Using FCIQMC-in-DFT, we establish here a much deeper understanding of the quantum states in NV centers in diamond, highlighting their multi-configurational nature and the consequent impact on the transition processes between different quantum states.

\section{Results}
\subsection{Model of NV Centers and Computational Framework}

The general strategy of FCIQMC-in-DFT calculations is overviewed in Fig. \ref{Fig_flowchart}.
An NV center is a point defect in diamond (left panel in Fig. \ref{Fig_flowchart}(a)), consisting of a substitutional nitrogen and a vacancy at its nearest neighbor site with $C_{3v}$ symmetry. 
The minimum model to describe the electronic structure of NV centers is the 4-orbital model.
As an example, right panel in Fig. \ref{Fig_flowchart}(a) shows a sketch of the single-particle picture of the negatively charged NV center (NV$^{-}$), which is the most stable state.
The electronic ground state of NV$^{-}$ includes the degenerate $e_{x,y}$ and the $a_{1}$ defect levels, i.e. one-electron state, in the conduction-valence band gap (5.4 eV).
In addition, an $a_{1}'$ orbital is within the valence band, near its top.
The arrows show the occupation of the six electrons in the four orbitals of the ground state of NV$^{-}$ \cite{PhysRevLett.77.3041}. 
To distinguish from the states within the single-configurational picture (e.g. $^{1}E$), we label the multi-configurational electronic states with $^{1}\bar{E}$. 
For NV$^{-}$, the most relevant electronic states are the four low-lying states shown in Fig. \ref{Fig_flowchart}(d), including the spin triplets ground state $^{3}\bar{A}_{2}$ and an optical active excited state $^{3}\bar{E}$, along with the intermediate spin singlets $^{1}\bar{A}_{1}$ and $^{1}\bar{E}$ \cite{gali_kaxiras_2008,larsson_electronic_2008}.
\begin{table}
\renewcommand\arraystretch{1.2}
\caption{The excitation energies (eV) of $^{1}\bar{E}\leftrightarrow{^{3}\bar{A}_{2}}$ and $^{3}\bar{E}\leftrightarrow{^{1}\bar{A}_{1}}$ in the negatively charged nitrogen vacancy (NV$^{-}$) obtained with different methods.}
\begin{ruledtabular}
\begin{tabular}{ccc}
 &\multicolumn{2}{c}{Excitation (eV)}\\
 Approaches&$^{1}\bar{E}\leftrightarrow{^{3}\bar{A}_{2}}$&$^{3}\bar{E}\leftrightarrow{^{1}\bar{A}_{1}}$\\
 \hline
 FCIQMC-in-DFT& 0.583&0.407\footnote{Combined with expt. in Ref. \cite{davies1976optical,rogers2008infrared}.} \\
 stochastic CI\footnote{Cluster results in Ref. \cite{delaney2010spin}.}&0.629, 0.644& \\
 G$_{0}$W$_{0}$-BSE\footnote{Ref. \cite{ma2010excited}.}&0.40&1.10 \\
 clust. Hub.\footnote{An extended Hubbard cluster model of Ref. \cite{ranjbar2011many}.}&0.41 &1.35\\
 CI / CI-RRA / CI-CRPA\footnote{Ref. \cite{bockstedte2018ab}.}&0.7 / 0.476 / 0.47&$\sim$0.1 / 0.545 / 0.61\\
 Quantum computer simulation\footnote{Ref. \cite{ma2020quantum,PRXQuantum.3.010339}.}&0.470 $\sim$ 0.561&0.243 $\sim$ 0.682\\
 ext. Hub. fit to GW\footnote{An extended Hubbard model fitted to GW in Ref. \cite{choi_mechanism}.}&$\sim$ 0.5&$\sim$ 0.5\\
 Expt.& &(0.344 $\sim$ 0.430)\footnote{Estimated by Ref. \cite{PhysRevLett.114.145502} with a model for intersystem crossing.}\\
\end{tabular}
\end{ruledtabular}
\label{NV_vertical}
\end{table}

Before the FCIQMC-in-DFT calculation, non-embedded calculations are performed to systematically determine the suitable settings, such as the size of the embedded region and environment; the exchange correlation functional of DFT; and the basis set.
Inspired by these calculations, we construct a cluster system C$_{42}$H$_{42}$N$^{-}$ and partition the total system C$_{42}$H$_{42}$N$^{-}$ into the "embedded region" C$_{3}$N$^{-}$ and the "environment" C$_{39}$H$_{42}$ (Fig. \ref{Fig_flowchart}b).
With the selected cluster, we start from a common non-local embedding potential constructed by DMFET, under which the sum of the subsystem DFT density matrices reproduces the reference density-matrix of the total system.
We follow sequentially a standard DMFET, an orbital orthogonalization, and a level-shifting projection term devised by Manby et al \cite{Manby2012}, and 
finally we get the embedding potential to perform subsequent FCIQMC-in-DFT calculations, which can describe the interaction effects of the environment on the embedded region (Fig. \ref{Fig_flowchart}c).
%
%To take into account the geometry relaxation effects, we carry out the embedding potential on the relaxed ground-state equilibrium structure (GES) with distorted $C_{s}$ symmetry.
%
%The subsequent FCIQMC calculations were performed with the RHF orbitals of "embedded region" C$_{3}$N$^{-}$ and this DFT embedding potential $V_{emb}^{A}$.
More details on embedding are described in the SI.
%
%Finally, the energy and configuration population of various states can be computed from FCIQMC, where quantities describing vertical excitation energy, intersystem crossing gap, and other charge states can be deduced.
%

\subsection{Negatively Charged NV Center}

\begin{figure*}
    \includegraphics[width=16.0cm]{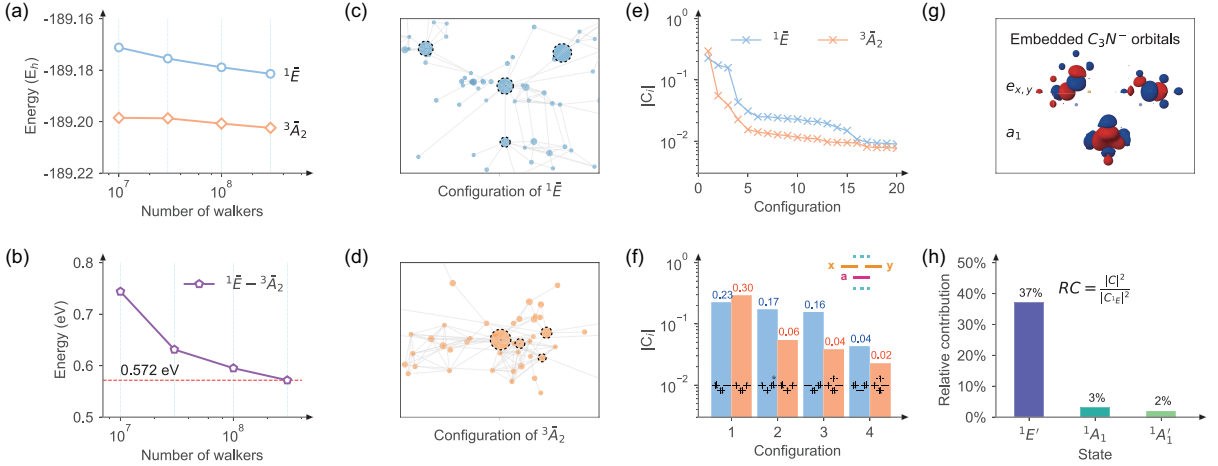}
    \caption{\textbf{FCIQMC calculations of NV$^-$.} (\textbf{a}) The FCIQMC energies of $^{1}\bar{E}$ and $^{3}\bar{A}_{2}$ as a function of the number of walkers. 
    (\textbf{b}) The corresponding excitation energies of $^{1}\bar{E}\leftrightarrow{^{3}\bar{A}_{2}}$. The red dashed line shows the final estimate at 0.572 eV. 
    (\textbf{c,d}) The configuration graphs of $^{1}\bar{E}$ and $^{3}\bar{A}_{2}$ states, characterizing the population and connection of configurations sampled from FCIQMC \cite{sun_ci_fig}. The size of circle represents the absolute value of CI coefficient, and the length of connecting lines correlates inversely with the transition matrix element between two configurations. The 4 most populated configurations have been highlighted by dashed lines, same as  (\textbf{f}).
    (\textbf{e}) The amplitude of the 20 most populated configurations from FCIQMC, same as (\textbf{c,d}).
    (\textbf{f}) The amplitude of the 4 most populated configurations of the spin triplet $^{3}\bar{A}_{2}$ (orange) and the singlet state $^{1}\bar{E}$ (blue). The corresponding configurations are described by frontier orbitals $x$ ,$y$ and $a$ (solid lines in inset, same as (\textbf{g})), as well as orbitals with lower energies and higher energies (dashed line in inset), occupied by 4 spin electrons. The star marker represents a normalized conjugate configuration. 
    (\textbf{g}) Embedded frontier molecular orbitals $e_{x,y}$ and $a_{1}$. 
    (\textbf{h}) In multi-configurational electronic state $^{1}\bar{E}$, the relative contributions (RC) of three mixed-in single-configurational states $^{1}E'$, $^{1}A_{1}$ and $^{1}A_{1}'$ with respect to the dominant single-configurational state $^{1}E$. In the formula of RC, $C$ represents the CI expansion coefficient corresponding to the single-configurational state.}
    \label{Fig_NV-}
\end{figure*}
%
%With FCIQMC-in-DFT calculations, we first discuss the singlet and triplet states of NV$^-$.
%
In Fig. \ref{Fig_NV-}(a-b) we plot the energies of $^{1}\bar{E}$, $^{3}\bar{A}_{2}$ of NV$^-$ and excitation energy of $^{1}\bar{E}\leftrightarrow{^{3}\bar{A}_{2}}$, as a function of the total number of walkers in FCIQMC.
The vertical excitation energy of $^{1}\bar{E}\leftrightarrow{^{3}\bar{A}_{2}}$ reaches 0.572 eV at $3\times10^{8}$ walkers (red dashed line), within an error of 0.01 eV.
%
%To further test whether our result is affected by the necessary approximations introduced, we consider several additional corrections based on a range of other calculations (Supplementary Note III).
%
Using the non-embedding calculations, we can further estimate the finite size effect of - 0.014 eV, the basis-set effect of - 0.004 eV, and the MOs truncation error of + 0.029 eV (SI. V).
%
%Although these error corrections are only rough estimate, we find that (i) they are quite small; (ii) they largely cancel with each other.
%
All the error corrections above only add up to + 0.011 eV, which makes very minor changes to our final results.
We can then correct the vertical excitation energy to 0.583 eV for $^{1}\bar{E}\leftrightarrow{^{3}\bar{A}_{2}}$ with embedded FCIQMC calculations.
In Table \ref{NV_vertical}, we compare the results with other state-of-the-art electronic structure methods.
For $^{1}\bar{E}\leftrightarrow{^{3}\bar{A}_{2}}$ excitation, 
our calculations agree quantitatively well with all the listed results, with the largest deviation of 0.2 eV.
%
%However, when a higher accuracy needs to be discussed, 
In particular, our result is consistent with the stochastic CI calculations of Delaney et al. \cite{delaney2010spin} as well as CI result of M. Bockstedte et al. \cite{bockstedte2018ab}, and the quantum computer simulation results of Ma et al. \cite{ma2020quantum}.
More precisely, our result is in between their two results with the maximum difference of 0.06 eV, within the so-called chemical accuracy.
%
%Ma et al. have also presented RPA calculations where the corresponding excitation energy is slightly underestimated further by $\sim$ 0.1 eV. 
%
%Similar or slightly more severe underestimations can be seen in the G$_0$W$_0$-BSE, CI-cRPA and an extended Hubbard cluster model calculations in previous reports.
%
%
%Table \ref{NV_vertical} also shows the excitation energy of $^{3}\bar{E}\leftrightarrow^{1}\bar{A}_{1}$, which can be obtained by combining the result of $^{1}\bar{E}\leftrightarrow{^{3}\bar{A}_{2}}$ of this study with the very reliable experimental measurements for the excitation energies between the two singlets and the two triplets.
%
Using the calculated vertical excitation energy 0.583 eV of $^{1}\bar{E}\leftrightarrow{^{3}\bar{A}_{2}}$, with the experimental
reference vertical excitation energy 2.18 eV \cite{davies1976optical} of $^{3}\bar{E}\leftrightarrow{^{3}\bar{A}_{2}}$ and excitation energy 1.19 eV \cite{rogers2008infrared} of $^{1}\bar{A}_{1}\leftrightarrow{^{1}\bar{E}}$, we also estimate that the excitation energy of $^{3}\bar{E}\leftrightarrow{^{1}\bar{A}_{1}}$ is 0.407 eV.
For this excitation, most other approaches have some degree of overestimation and vary greatly from one to another.
Goldman et al. \cite{PhysRevLett.114.145502,gali_ab} have concluded a region from 0.344 eV to 0.430 eV, through experimental measurement combined with a model, which is in good agreement with our result.

In addition to the excitation energies,
more importantly, FCIQMC calculations allow us to establish, without a priori assumption, the composition of the many-electron wavefunction of quantum states, which is expressed as $\Psi=\sum_{i} C_i \times D_i$, where $D_i$ is a configuration and $C_i$ is corresponding expansion coefficient.
Fig. \ref{Fig_flowchart}(a) presents the occupation of the reference determinant of the ground state $^{3}\bar{A}_{2}$ of NV$^{-}$.
The relevant frontier orbitals, $e_{x,y}$ and $a_{1}$, are further plotted in Fig. \ref{Fig_NV-}(g).
%
%
%The wavefunction of $^{3}\bar{A}_{2}$ is expressed linearly by all possible configurations $D_i$.
%
%\begin{equation}
%    \Psi= \sum_{i} C_i \times D_i
%\end{equation}
%$D_i$ is a Slater determinant of a given occupation form of orbitals, and $C_i$ is the so-called configuration interaction (CI) expansion coefficient.
%
%
%Noting that for a system consists of 30 electrons and 130 orbitals the number of possible configuration is gigantically $10^{40}$.
%
%With FCIQMC the contribution of all configurations can be sampled.
%
Fig. \ref{Fig_NV-}(c-d) shows the configuration graphs of the $^{1}\bar{E}$ and $^{3}\bar{A}_{2}$ states, following the scheme proposed in Ref. \cite{sun_ci_fig}. In the graph, each circle represents a configuration, and the lines characterize the connection between configurations. 
The size is proportional to the absolute value of the CI coefficient. 
A shorter line corresponds to a larger transition matrix element, along which FCIQMC spawning is more likely to happen.
We can see that both the triplet and singlet states are very multi-configurational, where the leading four configurations are disconnected from each other.
Fig. \ref{Fig_NV-}(e) plots the population of the top 20 configurations for both the $^{3}\bar{A}_{2}$ and $^{1}\bar{E}$ of NV$^{-}$.
The occupation of the leading configuration is less than 0.3, whereas in a typical system that is not strongly-correlated the leading occupation is often larger than 0.9.

The multi-configurational nature highlights the importance of electron correlations in NV centers.
Firstly, it is essential to the ISC transition, which can be observed in experiments.
%It was pointed out in literature that the correlations between different configurations are essential to the ISC transition, which is otherwise forbidden under the $C_{3v}$ symmetry of NV center.
%
%This can be explained by the occupations of electrons on frontier orbitals of several dominant configurations in Fig. \ref{Fig_NV-}(f).
%
%Specifically, when multi-configurational correlation exists, other states including $^{1}E'$ and $^{1}A_{1}$ can mix into $^{1}\bar{E}$, which can be correlated with $^{3}\bar{A}_{2}$, activating the ISC process.
%
We use $x$, $y$ and $a$ to refer to the corresponding configuration and for the ground state $^{3}\bar{A}_{2}$ the most dominant configuration is $\left | a\bar{a }xy  \right \rangle$.
The second most populated configuration is $\left | a\bar{x}xy  \right \rangle$, which accounts for $2\%$ of $\left | a\bar{a }xy  \right \rangle$.
Due to orbital degeneracy, single-configurational state $^{1}E$ is also doubly degenerate, written as $^{1}E_{x,y}$ \cite{Doherty_2011}. 
$^{1}E_{x,y}$ and $^{1}A_{1}$ are expressed as follows.
\begin{equation}
\begin{aligned}
    \left | ^{1}E_{x}  \right \rangle &=\frac{1}{\sqrt{2} } \left ( \left | a\bar{a }x\bar{x }   \right \rangle -  \left | a\bar{a }y\bar{y }   \right \rangle  \right )\\
    \left | ^{1}E_{y}  \right \rangle &=\frac{1}{\sqrt{2} } \left ( \left | a\bar{a }x\bar{y }   \right \rangle -  \left | a\bar{a }\bar{x }y   \right \rangle  \right )\\
    \left | ^{1}A_{1}  \right \rangle &=\frac{1}{\sqrt{2} } \left ( \left | a\bar{a }x\bar{x }   \right \rangle +  \left | a\bar{a }y\bar{y }   \right \rangle  \right )
\end{aligned}
\label{1E_1A1}
\end{equation}
In addition, the high energy states with $a$ orbital not been fully occupied are also relevant, including $^{1}E_{x,y}'$ and $^{1}A_{1}'$:
\begin{equation}
\begin{aligned}
    \left | ^{1}E_{x}'  \right \rangle &=\frac{1}{\sqrt{2} } \left ( \left | a\bar{x }y\bar{y }   \right \rangle -  \left | \bar{a }xy\bar{y }   \right \rangle  \right )\\
    \left | ^{1}E_{y}'  \right \rangle &=\frac{1}{\sqrt{2} } \left ( \left | a\bar{x }x\bar{y }   \right \rangle -  \left | \bar{a }x\bar{x }y   \right \rangle  \right )\\
    \left | ^{1}A_{1}'  \right \rangle
    &=\left | x\bar{x}y\bar{y}  \right \rangle\\
\end{aligned}
\label{1E'_1A1'}
\end{equation}
Combining the configuration population in Fig. 2(f) and Eq. \ref{1E_1A1}-\ref{1E'_1A1'}, the redefined mixed multi-determinant singlet state $^{1}\bar{E}$ is a mixture of single-configurational states $^{1}E$, $^{1}E'$, $^{1}A_{1}$ and $^{1}A_{1}'$.
Although $^{3}A_{2}$ is not linked to $^{1}E$ \cite{Doherty_2011}, but it is linked to $^{1}E'$ and $^{1}A_{1}$ through spin-orbit coupling (SOC).
So the multi-configurational state allows the transition that is otherwise forbidden in the single-configurational picture.
%which explains the experimental observations of ISC decay processes between $^{1}\bar{E}$ and $^{3}\bar{A}_{2}$.
%
Moreover, according to the previous analyses, our results suggest the ratio of ISC rates from $^{1}\tilde{E}$ toward $m_{s} = 0$ and $m_{s} =\pm 1$ of $^{3}\bar{A}_{2}$ is likely to decrease \cite{PhysRevB.98.085207}.

%Meanwhile, we notice that this minor contribution of $^{1}A_{1}$ is corresponding to pseudo-Jahn-Teller (PJT) effect, which is crucial for coupling two singlets $^{1}A_{1}$ and $^{1}\bar{E}$, as well as the temperature-dependent lifetime measurements of vibronic state $^{1}\tilde{E}$.
%
%The contribution of $^{1}E'$ in $^{1}\bar{E}$ to the dominant $^{1}E$ accounts for $36\%$.
%
Furthermore, it has been demonstrated that $^{1}E'$ can bring the dynamic Jahn-Teller (DJT) effect into $^{1}\bar{E}$ with a damped factor 0.01 \cite{PhysRevB.98.085207}, resulting
in a split of the E-vibronic states in the photoluminescence spectrum of the singlets \cite{Rogers_2008}.
Previous theoretical analyses have discussed the DJT effect by assuming that the correlated electronic state $^{1}\bar{E}$ consists of $^{1}E$ and $^{1}E'$ \cite{PhysRevB.98.085207}.
However, the relative contribution of $^{1}E'$ ($10\sim15\%$ in Ref. \cite{bockstedte2018ab,PhysRevB.98.085207}) to the dominant $^{1}E$ in $^{1}\bar{E}$ has been greatly underestimated.
We estimate, as Fig. \ref{Fig_NV-}(h) shows, the contribution of $^{1}E'$ is $37\%$.
When considering the relative contribution of $^{1}E'$ to $^{1}E$ in these two-state components model, we would get a larger DJT damped factor of 0.07.
This implies a stronger DJT effect and explains the large splitting energies of the two E-vibronic states in experiments.
%are higher than in previous theoretical analyses.
%
Besides DJT effect, pseudo-Jahn-Teller (PJT) effect contributes to the electron-phonon interactions of the shelving singlet states in NV$^{-}$, which couple the $^{1}\bar{A}_{1}$ and $^{1}\bar{E}$.
%with large energy gap.
%
In contrast to the enhanced contribution of DJT, our results indicate PJT effect would be relatively less important than the previous prediction \cite{PhysRevB.98.085207}.
We also notice the small relative contribution of $^{1}A_{1}$ ($3\%$) in $^{1}\bar{E}$, which is consistent with previous theoretical analyses \cite{gali_ab}, along with $2\%$ contribution of $^{1}A_{1}'$.
We expect the mixing of these high energy states to induce new electron-phonon interactions that have not been considered in current theoretical models.
%and require further investigation.
%
Overall, the multi-configurational nature of NV$^{-}$ states can be insightful for understanding the optical measurements, the electron-phonon interactions, and the spin selection mechanism in ISC, which are useful to improve spin initialization and readout in experiments.

\subsection{Neutral and Positively Charged NV Centers}

%
%So far, we have discussed the NV$^{-}$ being widely studied and applied, but NV defects include NV$^{0}$ and NV$^{+}$ charge states. 
%
%It has been demonstrated that photo-induced switching can lead to other charge states of NV centers in experiments, including the neutral NV$^{0}$ and the positively charged NV$^{+}$ \cite{pezzagna_quantum_2021}.
%
%In particular, the NV$^{+}$ charged state has shown promising properties for storage of nuclear spin qubit coherence, attributed to the absence of local electronic spin polarization \cite{pfender_protecting_2017}.
%

\begin{figure*}
    \includegraphics[width=16.0cm]{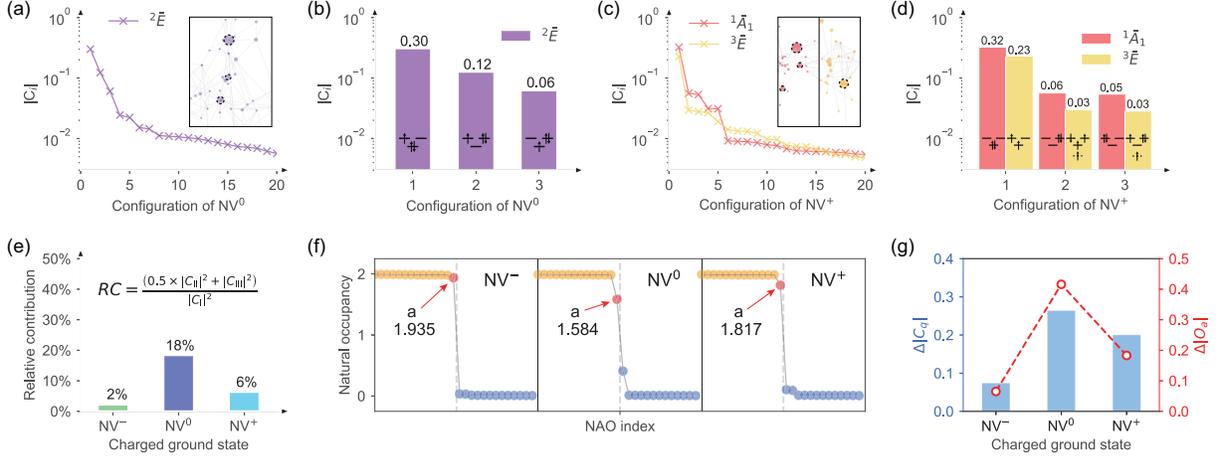}
    \caption{\textbf{Calculations on different charge states of NV centers.} 
    (\textbf{a}) Amplitude of the 20 most populated configurations for the ground state $^{2}\bar{E}$ of NV$^{0}$. The inset shows the corresponding configuration graph. 
    (\textbf{b}) The 3 dominating configurations for $^{2}\bar{E}$ of NV$^{0}$ with orbital occupation indicated. 
    (\textbf{c,d}) Same as (\textbf{a,b}) for the ground state $^{1}\bar{A}_{1}$ and excited state $^{3}\bar{E}$ of NV$^{+}$.
    %(\textbf{c}) Amplitude of the 20 most populated configurations for the ground state $^{1}\bar{A}_{1}$ and excited state $^{3}\bar{E}$ of NV$^{+}$. (\textbf{d}) The 3 dominating configurations for $^{1}\bar{A}_{1}$ and $^{3}\bar{E}$ of NV$^{+}$. 
    (\textbf{e}) For the ground states $^{3}\bar{A}_{2}$, $^{2}\bar{E}$ and $^{1}\bar{A}_{1}$ of the three charged NV centers, the relative contribution of the sum of the Type II and Type III configurations with respect to the Type I's (the first 3 most populated configurations only involving the $x$, $y$, $a$ orbitals). The coefficient 0.5 in the definition of relative contribution is resulted from the assumption that the depletion contribution of Type II is half that of Type III.
    (\textbf{f}) Natural occupancy for the ground states of the three charged NV centers. Red points highlight the natural orbital (NAO) occupancy corresponding to the $a$ orbital.
    (\textbf{g}) Blue histogram (left axis) shows difference between the experimental and DFT calculated values for nuclear spin quadrupole splittings ($\Delta|C_{q}|$ in unit of MHz) via data reported in Ref. \cite{pfender_protecting_2017}. Red line (right axis) shows the natural occupancy depletion of $a$ ($\Delta|O_{a}|$) calculated from FCIQMC.}
    \label{Fig_NV0_NV+}
\end{figure*}

Using the same embedding potential, we also perform the FCIQMC-in-DFT calculations for NV$^{0}$ and NV$^{+}$.
%
%Considering the same $C_{3v}$ symmetry representation of NV$^{-}$, the ground states of NV$^{0}$ and NV$^{+}$ are $^{2}\bar{E}$ and $^{1}\bar{A}_{1}$, respectively.
%
Fig. \ref{Fig_NV0_NV+}(a-d) plot the most populated configurations of ground states $^{2}\bar{E}$ and $^{1}\bar{A}_{1}$, as well as a spin triplet state of NV$^{+}$ labeled as $^{3}\bar{E}$.
The multi-configurational nature not only affects the electronic states, but also the charge density distribution in the vicinity of the NV defect.
Ref. \cite{pfender_protecting_2017} reported the experimental and DFT values for nuclear spin quadrupole splitting $C_{q}=3eQ_{N}V_{zz}/4h$ of $^{14}N$ for three NV centers at their ground states.
The splitting is proportional to the electric field gradient $V_{zz}$ at the nitrogen nucleus and the nuclear electric quadrupole moment $Q_{N}$.
From NV$^{-}$ via NV$^{0}$ to NV$^{+}$, electrons are preferentially excited from the $e_{x,y}$ orbitals, and the remaining electrons tend to increase their distance to the nitrogen nucleus, which leads to the reduction of $V_{zz}$, and consequently the decrease of the absolute value of quadrupole splitting $|C_{q}|$.
The experimental value of $|C_{q}|$ for NV$^{-}$, NV$^{0}$ and NV$^{+}$ are 4.945 MHz, 4.655 MHz and 4.619 MHz, respectively.
However, DFT calculation predicted a linear decrease of $|C_{q}|$ from NV$^{-}$ to NV$^{0}$ to NV$^{+}$, which is inconsistent with experiments.
The experimental values 
indicate that the charge density distribution at the vicinity of nitrogen of NV$^{0}$ is closer to NV$^{+}$ than NV$^{-}$.
In addition, as Fig. \ref{Fig_NV0_NV+}(g) shows, $|C_{q}|$ of each charge state is higher than the experiment: 0.075 MHz higher for NV$^{-}$,  0.265 MHz higher for NV$^{0}$, and 0.201 MHz higher for NV$^{+}$.
These differences can also be attributed to the multi-configurational nature of the corresponding states.
%Such differences indicate that single-configurational approximations overestimate the electric field at the nitrogen nucleus.
%
%They suggested that the correlated many-body state $^{2}\bar{E}$ cannot be accurate described by DFT.
%
%FCIQMC-in-DFT calculations show there is a $4\%$ contribution of $a(1)e(2)$ configuration to $^{2}\bar{E}$.
%which is consistent with our FCIQMC results of relative contribution $4\%$.
%
%Since only the $a_{1}$ orbital is localized on the nitrogen atom, as Fig. \ref{Fig_NV-}(g) shows,$a(1)e(2)$ configuration causes charge density depletion in the vicinity of nitrogen, thus the decrease of $V_{zz}$ and $|C_{q}|$.
%

%

%However, this cannot fully explain the overestimation of $|C_{q}|$ in the ground states of all three charge states and the similarity between NV$^{0}$ and NV$^{+}$.
%

%
Fig. \ref{Fig_NV0_NV+}(a-d) shows the configuration occupation obtained by FCIQMC-in-DFT for states of NV$^{0}$ and NV$^{+}$.
The ground states of the three charge states are $^{3}\bar{A}_{2}$ (NV$^{-}$), $^{2}\bar{E}$ (NV$^{0}$) and $^{1}\bar{A}_{1}$ (NV$^{+}$).
The dominant configurations of the three states all have $a_{1}$ orbital fully occupied, namely $\left | a\bar{a}yy  \right \rangle$, $\left | a\bar{a}x  \right \rangle$ and $\left | a\bar{a} \right \rangle$, respectively.
%
%These configurations are consistent with the single-configurational picture.
%
The non-negligible components where the $a_{1}$ orbital is singly occupied or empty cause delocalization of the charge density distribution around nitrogen atoms and the reduction of $|C_{q}|$.
We can divide the relevant configurations into three types: (I) $a_{1}$ orbital is fully occupied; (II) $a_{1}$ orbital is singly occupied; (III) $a_{1}$ orbital is unoccupied.
Then we can compute the relative contribution of the type II and type III configurations (RC defined in Fig. \ref{Fig_NV0_NV+}e),
%
%Since the amplitudes of the most populated components (Type I) of the three ground states are almost the same (0.3), we can compare the relative contributions.
%
%Because of the differences in the occupancy of $a_{1}$, the degree of charge density depletion of Type II shall be half of Type III, which contributes a factor of 0.5 in the definition of relative contribution.
%As shown in Fig. \ref{Fig_NV0_NV+}(e),
where we find the ground state $^{3}\bar{A}_{2}$ of NV$^{-}$ has a relative contribution of $2\%$, whereas $^{2}\bar{E}$ of NV$^{0}$ has $18\%$ and $^{1}\bar{A}_{1}$ of NV$^{+}$ has $6\%$.
%
%Considering the contributions' difference between a singly-occupied and empty $a_{1}$ orbital, the true value of Type II contributions may change further but this is a reasonable and reliable estimate.
%
The relative contribution qualitatively explains the trend of the depletion of charge density distribution at nitrogen atom, which is measured by $|C_{q}|$ in Fig. \ref{Fig_NV0_NV+}(e).
%The multi-configurational picture  linear trend of $|C_{q}|$ predicted by DFT \cite{pfender_protecting_2017}.
%
%It is obvious that the depletion in NV$^{0}$ is the highest and most influential, which results in the DFT value with 0.19 correction\cite{pfender_protecting_2017} still cannot describe $|C_{q}|$ accurately.
%

%
We also calculate the natural orbital (NAO) occupancy of the $a$ type orbital of the three ground states (Fig. \ref{Fig_NV0_NV+}(f)).
%
%We obtained additional insight from the single-particle density matrices calculated by embedded FCIQMC.
%
%The eigenvalues (the natural occupancy) and eigenvectors (NAOs)  highlight the important charge distribution near the Fermi level.
%
%As Fig. \ref{Fig_NV0_NV+}(f) shows, the natural occupancy in $a$ performs intuitive and significant differences among the three charge states.
%
For $^{3}A_{2}$ of NV$^{-}$, the natural occupancy is 1.935, in which the tiny depletion (compared to 2) results in a small decrease in $|C_{q}|$. 
%
%While for NV$^{0}$ and NV$^{+}$, things become much different.
%
The natural occupancy decreases to 1.584 for NV$^{0}$, indicating that the ionization process from NV$^{-}$ to NV$^{0}$ significantly affects the electric field near the nitrogen vicinity.
For NV$^{+}$, the natural occupancy is 1.818, and its charge distribution depletion degree is between NV$^{-}$ and NV$^{0}$.
Fig. \ref{Fig_NV0_NV+}(g) plots the natural occupancy depletion in $a$ orbital ($\Delta|O_{a}|$) for the three charged NV centers.
The trend is consistent with $\Delta|C_{q}|$, which is defined as the difference in experimentally measured and DFT calculated values of the nuclear spin quadrupole splitting \cite{pfender_protecting_2017}.
%
%
%We also note that the complete picture of the electronic structure and transition processes of the three charge states of NV centers is still not fully established.
%
%For example, the excited states of NV$^{+}$ are still lacking in studies, affecting our understanding of its optical processes.
%
%Based on the FCIQMC-in-DFT calculations, we predict that the first excited state is an spin triplet state $^{3}\bar{E}$, lying 1.523 eV above the ground state $^{1}\bar{A}_{1}$ in NV$^{+}$ (Fig. \ref{Fig_NV0_NV+}(d-e), Supplementary Note III ).
%
%The revealed multi-configurational nature may improve the understanding of the switching mechanism between charge states of NV defects.
%
%

%\section{Discussions}

\section{Conclusion}

To conclude, in this work we have integrated the FCIQMC methods with DMFET to achieve accurate \textit{ab initio} calculations of NV centers in diamond.
Our results provide a benchmark for the quantum states that are still a subject of uncertainty in experimental and theoretical studies.
The revealed multi-configurational nature of the quantum states in NV centers offers new insights to the understanding of sophisticated experiments in laboratory, which may further guide the optimization of photoelectronic quantum processes and lead to improved theoretical models. 
%
%
%As an outlook,FCIQMC, as a state-of-the-art electronic structure solver, is growing rapidly in terms of its capability to treat further shells of atoms around the defect \cite{Kai_NECI_2020,bogdanov_enhancement_2022}.
%
%New embedding schemes are also being proposed, which improve the convergence pattern when a larger cluster is needed \cite{shi_general_2022}, or when periodic supercells are used \cite{Knizia2013,masur_fragment-based_2016,nusspickel_systematic_2022,cui_systematic_2022}.
%
With continuous developments of FCIQMC and embedding theory, the computational framework of this study will allow us to tackle more complex quantum states in solids and defects in semiconductors, systematically towards the exact solution, providing valuable theoretical insights.

\begin{acknowledgments}
The authors thank Wei Fang for sharing a python wrap for geometry optimization.
This work was supported by the National Natural Science Foundation of China under Grant No. 92165101 and No. 11888101, and the Strategic Priority Research Program of Chinese
Academy of Sciences under Grant No. XDB33000000. 
We are grateful for computational resources provided by the TianHe-1A supercomputer, the High Performance Computing Platform of Peking University, the Platform for Data Driven Computational Materials Discovery of the Songshan Lake Materials Lab.
\end{acknowledgments}

\bibliography{ref.bib}

\end{document}